\documentclass[epj]{svjour}
\usepackage{graphics}
\usepackage{epsfig}
\begin{document}
\title{From Leading Hadron Suppression to Jet Quenching at RHIC and at the LHC}
\author{Urs Achim Wiedemann 
}                     
%
%
\institute{Department of Physics, CERN, Theory Division, CH-1211 Geneva 23}
\date{Received: date / Revised version: date}
%
\abstract{
In nucleus-nucleus collisions at the Relativistic Heavy Ion Collider (RHIC), 
one generically observes a strong medium-induced suppression of 
high-$p_T$ hadron production. This suppression
is accounted for in models which assume a significant medium-induced 
radiative energy loss of high-$p_T$ parent partons produced in the collision.
How can we further test the microscopic dynamics conjectured to underly
this abundant high-$p_T$ phenomenon? What can we learn about the dynamics
of parton fragmentation, and what can we learn about the properties of the
medium which modifies it ? Given that inelastic parton scattering is expected
to be the dominant source of partonic equilibration processes, can we use 
hard processes as an experimentally well-controlled window into QCD 
non-equilibrium dynamics ? Here I review what has been achieved so far,
and which novel opportunities open up with higher luminosity at RHIC, and
with the wider kinematical range accessible soon at the LHC.
} 
\maketitle
\section{Introduction}
\label{sec1}
High-$p_T$ partons, propagating through the dense and spatially extended QCD matter created
in a nucleus-nucleus collision, are expected~\cite{Baier:2000mf,Kovner:2003zj,Gyulassy:2003mc,Baier:1996sk,Zakharov:1997uu,Wiedemann:2000za,Gyulassy:2000er,Wang:2001if} to suffer a 
significant medium-induced energy degradation prior to hadronization in the vacuum. 
Similar to electric charges propagating through QED matter, inelastic radiative contributions 
dominate over elastic collisional ones~\cite{Baier:2000mf,Thoma:1995ju} for highly energetic projectiles. However, compared to its abelian analogue,
radiative QCD energy loss shows characteristic differences which can be traced back 
to its non-abelian nature. In contrast to QED photon radiation, the gluonic 
quanta radiated off the hard projectile are color charged, and their interaction with the 
medium is even stronger than that of the parent projectile. This implies a significant
$p_T$-broadening of the radiated remnants, and an approximately quadratic dependence 
of parton energy loss on in-medium path length.

The second qualitative difference between QCD parton energy loss and standard
electrodynamic energy loss processes is that typical high-$p_T$ partons are of 
high virtuality, $Q^2 \sim p_T^2$. Even in the absence of a nuclear environment, 
the uncertainty principle dictates that these partons must reduce their virtuality quickly
on time (length) scales which turn out to be comparable to a nuclear diameter. 
They do so by breaking up into remnants of reduced virtuality. In the vacuum, this 
perturbative fragmentation process is described by an angular ordered DGLAP parton 
shower which has a probabilistic interpretation, and which underlies analytical 
approaches and Monte Carlo simulations of parton fragmentation. In the presence 
of a medium, there is an interference pattern between this 
"vacuum radiation" and the medium-induced radiation~\cite{Wiedemann:2000za}. As a consequence,
the leading parton (and a fortiori the leading hadron) in the shower emerges with reduced 
energy, the associated multiplicity of the jet is increased and softened, and the shower 
becomes broader.  

Models based on this picture~\cite{Wang:2003aw,Dainese:2004te,Eskola:2004cr,Gyulassy:2000gk,Drees:2003zh,Vitev:2002pf,Hirano:2002sc} account for the main 
modifications of high-$p_T$ hadron production in nucleus-nucleus collisions 
at RHIC, namely the strong suppression of single inclusive hadron spectra, their centrality 
dependence~\cite{Adcox:2001jp,Adler:2003au,Adler:2002xw,Adams:2003kv,Back:2003qr,Arsene:2003yk}, the corresponding suppression of leading back-to-back 
correlations~\cite{Adler:2002tq}, and high-$p_T$ hadron production with respect 
to the reaction plane~\cite{Adams:2004wz}. The above picture of the microscopic dynamics of 
medium-induced parton energy loss also makes predictions for two classes of measurements 
which are now gradually coming into experimental reach~\cite{Wiedemann:2004wp}, namely
i) high-$p_T$ particle correlations~\cite{Majumder:2004wh,Majumder:2004pt}, jet shapes and jet multiplicity distributions~\cite{Salgado:2003rv,Pal:2003zf,Armesto:2004pt,Armesto:2004vz} which test the predicted relation between the energy loss of the leading 
parton, the transverse momentum broadening of the parton shower, and the softening of its 
multiplicity distribution and ii) the relative yields of identified high-$p_T$ hadrons 
which test the prediction that medium-induced parton energy loss depends on the color
charge and mass of  the parent parton. 

In my view, the main challenge for jet physics in nuclear collisions remains to gain a better understanding of the relation between the observed medium modifications, the partonic 
nature of equilibration processes and the properties of the produced matter~\cite{Wiedemann:2004wp}. Progress in this direction is most likely to occur via a detailed testing of the microscopic dynamics 
of parton energy loss in a close interplay of theory and experiment. This is made possible 
by two basic properties of hard hadronic probes: they are {\it abundant} and they are 
{\it strongly sensitive} to medium effects. 
Their sensitivity ensures that the experimentally measured medium-effects are larger 
than the estimated theoretical and experimental uncertainties, and their abundant
rates allow us to quantify more and more differential properties. Both features are prerequisits
for a detailed characterization of the medium-modified partonic fragmentation pattern.
This talk aims at emphasizing that such a detailed characterization is not
{\it l'art pour l'art}, but provides one of the most promising approaches to learn about
QCD equilibrium and non-equilibrium physics.

\section{Current theory of parton energy loss}
\label{sec2}

Parton propagation in a medium can be characterized by solving the Dirac
equation for the corresponding partonic wave function in the color field of the 
medium~\cite{Buchmuller:1995mr,Wiedemann:2000ez}. In the limit of high 
projectile energy, the resulting $S$-matrix of any incoming parton is
given by an eikonal path-ordered Wilson line
\begin{equation}
W({\bf x}) = {\cal P} \exp \left[i \int dx^- A_a^+({\bf x},z^-)\, T^a \right]\, . 
	\label{2.1}
\end{equation}
Here, $T^a$ is the $SU(3)$ generator in the representation of the propagating parton. 
The transverse coordinate ${\bf x}$ of the parton is frozen during propagation, but the 
parton rotates in color space due to the interaction with the target color field $A_a^+$ 
characterizing the medium. In general, production cross sections measure the extent to
which different virtual components of the incoming partonic wavefunction pick up
different phases and decohere during the scattering~\cite{Kovner:2001vi}. The target average of two
fundamental Wilson lines at different transverse coordinates,
\begin{equation}
	N({\bf x},{\bf y}) = 1 - 
	\langle {\bf Tr} \left[ W({\bf x})\, W({\bf y}) \right]  \rangle / N_c\, ,
	\label{2.2}
\end{equation}
is the simplest medium-dependent quantity which measures this phase decoherence.
The high-energy limit of a large class of processes, including diffractive 
processes\cite{Buchmuller:1995mr}, gluon production in
proton-nucleus collisions~\cite{Kovchegov:1998bi,Kovner:2001vi} and the photoabsorption 
cross section in the limit of small Bjorken $x$~\cite{Wiedemann:2000ez,Kovner:2001vi}, 
have been formulated such that the target average (\ref{2.2}) 
provides the entire information about the nuclear target.

In the study of medium-induced parton energy loss, we seek the multiple scattering-induced
modifications of the bremsstrahlung gluon spectrum
\begin{equation}
	dI^{\rm vac} = C_{F,A} \frac{\alpha_s}{\pi} \frac{dk^2_{\perp}}{k_\perp^2}
		                \frac{d\omega}{\omega}\, .
		                 \label{2.3}
\end{equation}
The total gluon bremsstrahlung spectrum is of the form
\begin{equation}
	\omega\frac{dI^{\rm tot}}{d\omega\, d{\bf k}_\perp}
	= \omega\frac{dI^{\rm vac}}{d\omega\, d{\bf k}_\perp} 
	+ \omega\frac{dI^{\rm med}}{d\omega\, d{\bf k}_\perp}\, ,
	\label{2.4}
\end{equation}
where the medium modifications are power-suppressed, but parametrically larger 
than other power-suppressed corrections due to geometrical enhancements proportional 
to the large nuclear size. The distance between the production point of the hard parton 
in space-time and the typical position of a secondary scattering introduces a microscopic
length scale (mean free path) relevant for the interference between vacuum and 
medium-induced radiation. To account for that, one has to go beyond the eikonal
approximation by including the leading transverse Brownian motion of the produced partons.
One finds that the medium-induced modification of (\ref{2.3}) depends on the medium
via the path-integral~\cite{Zakharov:1997uu,Wiedemann:2000za}
\begin{equation}
	{\cal K}({\bf r}_1,z_1;{\bf r}_2,z_2) =\int_{{\bf r}(z_1)}^{{\bf r}(z_2)} \hspace{-0.5cm}
	 {\cal D}{\bf r} \exp\left[ \int_{z_1}^{z_2} \hspace{-0.3cm} d\xi
	 \left[ i\frac{\omega}{2} \dot{\bf r}^2 - \frac{1}{2} n(\xi)    
	\sigma(\bf r) \right]\right]\, ,
	\label{2.5}
\end{equation}
where
\begin{equation}
	 e^{- \frac{1}{2} n(\xi) \sigma(\bf r) } \equiv \Big\langle  e^{i  A_a^+({\bf x},\xi)\, T^a}  
	 e^{- i  A_b^+({\bf x}+{\bf r},\xi)\, T^b}
	 \Big\rangle \, .
	 \label{2.6}
\end{equation}
In the limit of large gluon energy $\omega$, the path integral (\ref{2.5}) approaches
the eikonal limit (\ref{2.2}) of the target average of two Wilson lines. The
expressions (\ref{2.2}) and (\ref{2.6}) encode essentially the same physical information, 
namely the transverse color field strength which the target presents to the
hard partonic projectile. In (\ref{2.6}), this information is parametrized in terms 
of a longitudinal density $n(\xi)$ times a measure of the transverse field strength per
unit pathlength, $\sigma({\bf r})$. In terms of this target average, the medium-modification
(\ref{2.4}) of the gluon bremsstrahlung spectrum to leading order in parton energy
and for arbitrarily large target color field reads~\cite{Wiedemann:2000za,Wiedemann:2000tf}
\begin{eqnarray}
  && \omega\frac{dI^{\rm med}}{d\omega\, d{\bf k}_\perp}
  = {\alpha_s\,  C_R\over (2\pi)^2\, \omega^2}\,
    2{\rm Re} \int_{\xi_0}^{\infty}\hspace{-0.3cm} dy_l
  \int_{y_l}^{\infty} \hspace{-0.3cm} d\bar{y}_l\,
   \int d{\bf u}\, e^{-i{\bf k}_\perp\cdot{\bf u}}   
  \nonumber \\
  && \qquad \times   
  e^{ - \frac{1}{2} \int_{\bar{y}_l}^{\infty} d\xi\, n(\xi)\, \sigma({\bf u}) }\,
  {\partial \over \partial {\bf y}}\cdot
  {\partial \over \partial {\bf u}}\,
  {\cal K}({\bf y}=0,y_l,{\bf u},\bar{y}_l)\, .
    \label{2.7}
\end{eqnarray}
For numerical applications, one approach is the opacity expansion of (\ref{2.7}) 
which amounts to expanding the integrand in powers of the density 
$n(\xi)$~\cite{Wiedemann:2000za,Gyulassy:2000er}. Alternatively,
one exploits that the main support of the integrand of (\ref{2.7}) is for small transverse
distances which permits  the harmonic oscillator approximation of the path integral with
~\cite{Zakharov:1997uu,Baier:1996sk,Wiedemann:2000za}
\begin{equation}
   n(\xi)\, \sigma({\bf r}) \approx \frac{1}{2} \hat{q}(\xi)\, {\bf r}^2\, .
   \label{2.8}
\end{equation} 
Here, $ \hat{q}$ is the so-called BDMPS transport coefficient which parametrizes the
small distance behavior of the expectation value of the gauge invariant operator in 
(\ref{2.2}) [see also (\ref{2.6})]. Physically, $ \hat{q}$ characterizes the average
squared transverse momentum transferred from the medium to a hard gluon per unit
path length. Remarkably, for a static medium without time evolution, the small-distance 
properties of the Wilson line average (\ref{2.6}) can be shown to determine the main 
properties of the medium-induced gluon radiation spectrum. Momentum broadening is 
characterized by $\langle k_\perp^2 \rangle \sim \hat{q} L_{\rm med}$~\cite{Baier:1996sk}, 
the energy distribution is determined by the characteristic  energy scale 
$\omega_c = \frac{1}{2} \hat{q} L_{\rm med}^2$~\cite{Baier:1996sk},  and medium effects 
regulate the additional gluon radiation in the infrared on 
a scale $\omega \sim \omega_c/\left( \hat{q} L_{\rm med}^3\right)^{2/3}$ ~\cite{Salgado:2003gb}.
For an expanding medium, the transport coefficient decreases with time - this translates
into a path dependence $\hat{q}=\hat{q}(\xi)$~\cite{Baier:1998yf,Gyulassy:2000gk,Salgado:2002cd}. 
One finds, however, that the medium-induced gluon radiation spectrum for a dynamically 
expanding case is the same as that obtained for a static medium of rescaled linear 
line-averaged transport coefficient~\cite{Salgado:2002cd}
\begin{equation}
  \bar{\hat{q}} = \frac{2}{L_{\rm med}^2} \int_0^{L_{\rm med}}
  \tau\, \hat{q}(\tau)\, d\tau\, .
  \label{2.9}
\end{equation}
In practice, this dynamical scaling law allows a simplified data analysis in terms of static 
medium properties; a posteriori, one then translates the extracted static 
average transport coefficient into the realistic dynamical one. In what follows, we often
denote the time-averaged transport coefficient by $\hat{q}$.

\section{Leading hadron spectra}
\label{sec3}

Up until now, experimental information about parton energy loss comes mainly from
the study of single inclusive hadron spectra at high transverse momentum. How do
we quantify "high" ? A qualitative assumption of the current picture of parton energy loss 
is that high-$p_T$ partons are sufficiently Lorentz boosted to hadronize at large length
scales (late times) outside the medium~\cite{Wiedemann:2004wp}. As a consequence, 
parton energy loss affects the {\it ratio} of identified hadron spectra at high $p_T$ only to the extent
to which the parent partons suffer different energy loss effects. Thus, parton energy loss can be 
the dominant medium effect only for $p_T > 6$ GeV, above the kinematical region in which 
the baryon-to-meson ratio shows an anomalous enhancement. 

The transverse momentum of the leading hadron traces the energy of the leading parton
in the parton shower and thus contains information about parton energy loss. However, the measurement of high-$p_T$ hadrons induces a significant bias on the partonic fragmentation. 
For example, in a light quark jet, the leading hadronic fragment carries on average $\sim 1/4$ 
of the total jet energy, but a typical hadron in the high-$p_T$ tail of a single inclusive spectrum
carries typically $\sim 3/4$ of the energy of its parent parton. In the presence of a medium, 
this so-called trigger bias effect is medium-dependent: the parton which contributes to the 
high-$p_T$ tail of the hadronic spectrum belongs to a fragmentation pattern which
is much harder than the average, and thus it has an in-medium path length which is
significantly smaller than the average. Also, for any given path length, it has a medium-induced
energy loss which is smaller than the average. This explains why the notion of {\it average}
parton energy loss is not suitable for calculating high-$p_T$ hadron suppression~\cite{Baier:2001yt}. 
What is needed in practice is the distribution in probability that a part $\Delta E$ 
of the total parton energy is radiated away. For the case of radiation of an arbitrary 
number of $n$ {\it independent} gluon emissions, this probability reads~\cite{Baier:2001yt,Salgado:2003gb}
\begin{eqnarray}
  && P(\Delta E, L, \hat{q}) = \sum_{n=0}^\infty \frac{1}{n!}
  \left[ \prod_{i=1}^n \int d\omega_i \frac{dI(\omega_i)}{d\omega}
    \right]
    \nonumber \\
    && \qquad \times 
    \delta\left(\Delta E - \sum_{i=1}^n \omega_i\right) 
    \exp\left[ - \int_0^\infty \hspace{-0.2cm}
      d\omega \frac{dI}{d\omega}\right]\, .
   \label{3.1}
\end{eqnarray}
To model the effects of parton energy loss, the quenching weights ${\cal P}$ are convoluted 
with hard partonic cross sections and the parton fragmentation functions~\cite{Eskola:2004cr}.
Since ${\cal P}$ is a probability distribution, this accounts for the above-mentioned trigger bias effects~\cite{Salgado:2003gb,Baier:2001yt}.  

\subsection{Determining the BDMPS transport coefficient $\hat{q}$ from $R_{AA}$ for
light-flavored hadrons}

The medium-induced suppression of single inclusive hadron spectra
$d^2N^{AA}/dp_Tdy$ in nucleus-nucleus ($AA$) collisions is commonly quantified 
in terms of the nuclear modification factor
\begin{equation}
   R_{AA}(p_T,y)=\frac{d^2N^{AA}/dp_Tdy}{\langle T_{AA}\rangle_c
                 d^2\sigma^{NN}/dp_Tdy}\, .
   \label{3.2}
\end{equation}
Here, $\langle T_{AA}\rangle_c$ is the standard nuclear overlap 
function, calculated as the average in the measured centrality class.
In the absence of nuclear or medium effects, $R_{AA}\equiv 1$.
We have calculated this factor for realistic nuclear geometry as described 
above~\cite{Eskola:2004cr}: For the hadron yield
in AA collisions, nuclear modifications of the parton distribution functions~\cite{Eskola:1998df}
and parton energy loss was taken into account. The reference yield for nucleon-nucleon
collisions is calculated in the same formalism without these medium effects. In the absence
of parton energy loss, $R_{AA}$ reduces slightly with increasing $p_T$ since the relevant 
values of Bjorken $x$ move from the anti-shadowing to the so-called
EMC region where nuclear parton distribution functions decrease
faster than proton ones with increasing $x$~\cite{Eskola:2002kv}. 
However, these pdf effects cannot account for the strength of the suppression. 
In contrast, final state parton energy loss can
account for the degree of suppression. 

\begin{figure}[h]
\begin{center}
\vspace{-1cm}
\epsfysize=7cm\epsffile{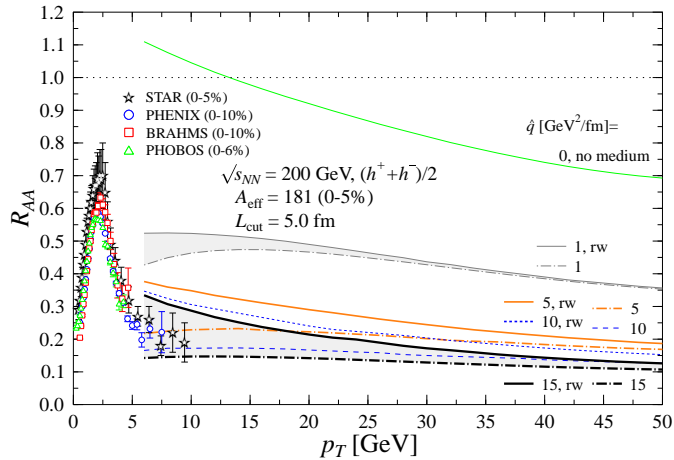}
\caption[a]{The nuclear modification factor $R_{AA}$ for charged
hadrons in the 0-5 \% most central Au+Au collisions at 
$\sqrt{s_{NN}} = 200$ GeV for different values of the time-averaged
transport coefficient $\hat q$. Differences between solid and
dash-dotted curves indicate uncertainties related to finite
energy corrections. Figure taken from~\cite{Eskola:2004cr}.
}
\label{fig1}
\end{center}
\end{figure}

Two features are remarkable~\cite{Eskola:2004cr}:
First, the calculation indicates an approximately $p_T$-independent
suppression pattern. The reason is that the slope of the partonic $p_T$-spectrum 
increases at RHIC gradually with increasing $p_T$. This is a trigger bias, according
to which for the same value of  $R_{AA}$, partons at higher $p_T$ have to lose a 
smaller fraction of their total energy. Indeed, close to the kinematical boundary
$p_T \sim O(\sqrt{s}/2$, for $p_T > 20-25$ GeV, the partonic spectrum at RHIC does 
not follow a power law but it has an approximately exponential shape.
In this limit, the $p_T$-independence of $R_{AA}$ is obvious, since
\begin{equation}
	R_{AA} = \frac{\int d\Delta E\, {\cal P}(\Delta E)\, 
		\exp\left[ -(p_T+\Delta E)/E_{\rm slope}  \right]}
		{\exp\left[ -(p_T)/E_{\rm slope}  \right]}\, . 
		\label{3.3}
\end{equation}
Secondly, Fig.~\ref{fig1} indicates that the absolute size of the suppression
$R_{AuAu}\sim 0.2$ does not require a fine tuning of parameters but appears
naturally from the interplay of parton energy loss and nuclear geometry. 
Indeed, as long as the density of the medium is sufficiently high (i.e. $\hat{q}$
is large enough), $R_{AuAu}$ approaches a factor 5 suppression for central
collisions. However, even if one increases the density further, the suppression
increases only slightly, since an essentially fixed fraction of the hard partons is 
produced in the outer corona of the two-dimensional transverse overlap of the
nucleus-nucleus collision and thus remains almost unaffected
due to a negligible in-medium path-length~\cite{Muller:2002fa,Dainese:2004te,Eskola:2004cr}. 
This also implies that the sensitivity of single inclusive spectra to properties of the
medium is rather limited: they provide only a lower bound on the produced density.

The model discussed above also accounts for the observed centrality dependence 
of the nuclear modification factor~\cite{Wang:2003aw,Wang:2003mm,Dainese:2004te}. 
However, information about the centrality dependence does not appear to increase the 
sensitivity to properties of the medium. In fact, on the basis of the measured centrality
dependence, one cannot even discriminate between a linear and a 
quadratic dependences of parton energy loss on $L_{\rm med}$~\cite{Drees:2003zh}. 

\subsection{Dependence of parton energy loss on parton identity}
The calculation of medium-induced gluon radiation predicts a hierarchy of
parton energy loss depending on parton identity,
\begin{equation}
	\Delta E_{\rm gluon} > \Delta E_{\rm quark,\, m=0} > \Delta E_{\rm quark,\, m\not=0} \, .
	\label{3.4}
\end{equation}
Gluons lose more energy than quarks due to their stronger coupling to the medium
({\it color charge effect}), and massive quarks lose even less energy than massless 
ones due to the mass-dependent phase-space restrictions on the medium-induced
gluon radiation spectrum ({\it mass effect})~\cite{Dokshitzer:2001zm,Armesto:2003jh,Djordjevic:2003zk,Zhang:2003wk}.
Testing the hierarchy (\ref{3.4}) is an important step towards verifying the microscopic
partonic dynamics conjectured to underly high-$p_T$ hadron suppression. 

\begin{figure*}[t]\epsfxsize=11.7cm
\centerline{\epsfbox{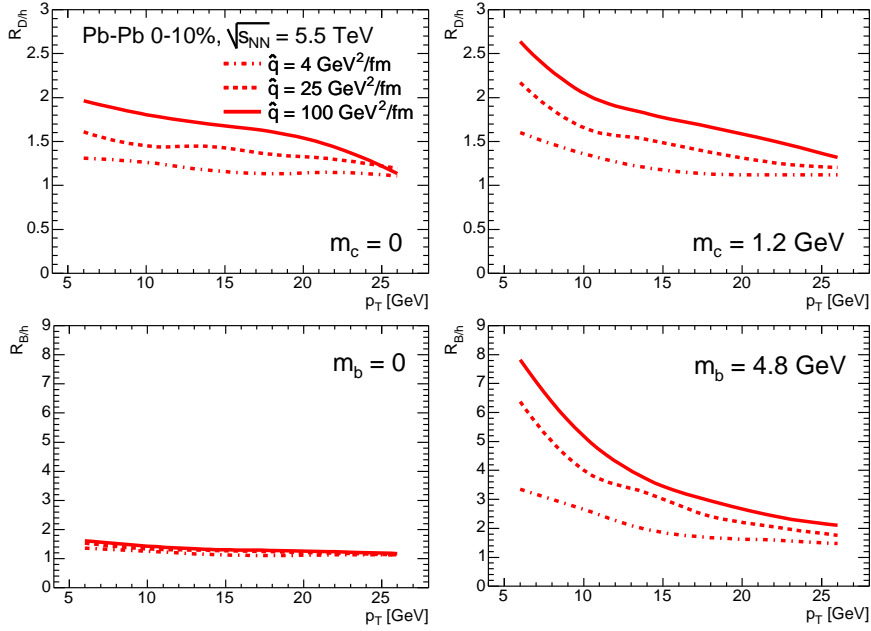}}
\vspace{0.5cm}
\caption{Heavy-to-light ratios for $D$ mesons (upper plots) and $B$ mesons (lower
plots) for the case of a realistic heavy quark mass (plots on the right) and for a case
study in which the quark mass dependence of parton energy loss is neglected
(plots on the left). Figure taken from \protect\cite{Armesto:2005iq}.
}\label{fig8}
\end{figure*}

On the level of single inclusive spectra, one can test (\ref{3.4}) by comparing the 
suppression of identified hadron spectra which differ in their parent partons. 
Of particular interest is the ratio of nuclear modification factors of high-$p_T$
heavy-flavored mesons to light-flavored hadrons (``heavy-to-light ratio'')~\cite{Armesto:2005iq}.
In general, these ratios are sensitive to the color charge and the mass effect, but
their sensitivity to both effects differs with center of mass energy, transverse momentum
range and hadron identity. In Fig.~\ref{fig8}, results are shown for the heavy-to-light
ratios of $D$ and $B$ mesons at the LHC. For $D$ meson spectra at high but experimentally 
accessible transverse momentum ($10\sim p_T \sim 20~{\rm GeV}$) in Pb--Pb collisions 
at the LHC, one finds that charm quarks behave essentially like light quarks. However, 
since light-flavored hadron yields are dominated by gluon parents, the
heavy-to-light ratio of $D$ mesons is a sensitive probe of the color charge
dependence of parton energy loss. In contrast, due to the larger $b$ quark mass, the
medium modification of $B$ mesons in the same kinematical regime provides a
sensitive test of the mass dependence of parton energy loss.  Hence, a combined
analysis of $D$ and $B$ meson spectra at LHC will allow to clarify the parton species
dependence of parton energy loss. At RHIC
energies, the strategies for identifying and disentangling the color charge
and mass dependence of parton energy loss are more involved. First, high-$p_T$
hadron spectra test parton distribution functions at $\sim 30$ times higher Bjorken
x than at the LHC. This implies that light-flavored hadrons have a more significant
contribution of parent quarks, and heavy-to-light ratios are thus less sensitive to
the color charge effect. Second, the kinematical regime suited for a characterization 
of the mass effect in with $D$ meson spectra at RHIC is limited to $7 \sim p_T \sim
12~{\rm GeV}$. On the other hand, in heavy-to-light ratios, color charge and mass effect 
do not compensate each other but lead both to an enhancement. So, even if it should 
be difficult to disentangle both effects, their combined contribution can be expected to
leave a valuable additional information about parton energy loss. 

Particle correlations may provide important complementary information for elucidating 
the influence of parton identity on final state parton energy loss.
For example, requiring that a high-$p_T$ trigger hadron at forward rapidity is balanced
by a recoil at mid-rapidity, one may be able to study medium-modified hadron production
in a configuration which enriches the contribution of gluon parents. Both at RHIC and at
the LHC, the class of correlation measurements with this potential is large. 
It is an open question whether some of these measurements have a similar 
or even higher sensitivity to the mass and color charge dependence of parton energy 
loss than the ratios of particle identified single inclusive hadron spectra discussed above.

\subsection{What do we learn from $R_{AA}$ about the medium ?}

The first answer to this question is a number: the experimentally favored value
of the time-averaged transport coefficient, $\overline{\hat q} \simeq 10\ {\rm GeV}^2/{\rm fm}$. 
Formally, this quantity characterizes the small-distance property of the expectation value 
(\ref{2.2}) of two light-like Wilson lines in the target average characterizing the produced 
medium. The main question is whether this value for $\overline{\hat q}$ is consistent with 
expectations for a thermalized system. For this problem, standard lattice techniques do not 
apply, and it is an open challenge of how to perform the calculation of the average (\ref{2.2}) 
in a thermal heat bath. Novel approaches are clearly needed, and may involve new concepts
such as exploiting AdS/CFT correspondence~\cite{Sin:2004yx}. On a less ambitious level,
even if the initial conditions of (\ref{2.2}) cannot be calculated, the energy dependence of 
(\ref{2.2}) should be predictable since it is expected to satisfy non-linear 
small-$x$ evolution equations. 

The only attempt so far to calculate the transport coefficient $\overline{\hat q}$ for a QCD 
equilibrium state is rather model dependent~\cite{Baier:2002tc} [for another recent discussion,
see also Ref.~\cite{Muller:2005en} in these proceedings]. In general, $\hat q$ is expected to
be proportional to the number density of scattering centers, and hence for an
ideal gas
\begin{equation}
  \hat q(\tau)= c\,  \epsilon^{3/4}(\tau)\, .
   \label{3.5}
\end{equation}
Here, $c$ is a medium-dependent proportionality constant. To determine it for a quark 
gluon plasma, Ref.~\cite{Baier:2002tc}
estimates the momentum transfer from the medium to the hard parton per 
medium constituent by modeling these constituents with temperature dependent 
Debye-screened scattering potentials.  One finds~\cite{Baier:2002tc} 
\begin{equation}
 c_{\rm QGP}^{\rm ideal} \approx 2\, .
 \label{3.6}
\end{equation}
This estimate is based on the strong assumptions spelled out above; clearly, an 
alternative calculation would be desirable. Here, we explore the
consequences if the number (\ref{3.6}) is taken seriously.

For an expanding medium with expansion parameter $\alpha$,
and $\hat q(\tau) = \hat q(\tau_0)\, \left(\tau_0/L 
\right)^\alpha$, one finds from the dynamical scaling law (\ref{2.9})
\begin{eqnarray}
  c &=& \frac{\hat q (\tau_0)}{\epsilon(\tau_0)^{3/4}}
    = \frac{\overline{\hat q}}{\epsilon(\tau_0)^{3/4}}
      \frac{2-\alpha}{2} 
  \left( \frac{L}{\tau_0} \right)^{\alpha}\, .
  \label{3.7}
\end{eqnarray}
For a typical initial time $\tau_0 \sim 0.2$ fm/c, the average in-medium
pathlength may be as small as $L \sim 10\, \tau_0 = 2$ fm .
The energy density averaged over a uniform transverse profile
at initial time $\tau_0$ can be as large as $\epsilon(\tau_0) \sim 
O (100) \frac{\rm GeV}{{\rm fm}^3}$~\cite{Eskola:1999fc}, although
much smaller estimates exist~\cite{Armesto:2004ud}
This implies that realistic scenarios lie in the
parameter range $\epsilon(\tau_0) < 100\, \frac{\rm GeV}{{\rm fm}^3}$,
$L\sim 10\, \tau_0$, $0.75 < \alpha < 1.5$, for which one finds
\begin{eqnarray}
   c> 8\ldots 19 > (4-5)\,  c_{\rm QGP}^{\rm ideal}\, .
   \label{3.8}
\end{eqnarray} 
Taken at face value, this number quantifies the extent to which the produced
medium deviates from an ideal quark gluon plasma. Interactions between the
medium and the hard partonic test particle are at least a factor $\sim 4 - 5$ 
stronger than perturbatively expected. Remarkably, to reproduce an elliptic flow 
in agreement with experiment, parton cascade 
calculations require  partonic cross sections which are a similar factor
$\sim 4 - 5$ larger than perturbatively expected~\cite{Molnar:2001ux}. 
This supports the picture that hard partons are {\it test particles} which participate
in the same equilibration processes as bulk degrees of freedom. It strengthens 
the motivation that the analysis of medium-modified hard parton fragmentation
provides access to QCD equilibration dynamics. This leads us to several
far-reaching questions all of which require a better dynamical understanding of
the transport coefficient $\hat{q}$. For example: Is an interaction strength of the 
size (\ref{3.8}) compatible with the known residual interactions in a QGP which 
result in a $\sim 20 \%$ deviation of the energy density from the ideal Stefan-Boltzmann 
limit at high temperature ?

\section{Medium-modification of jets}
\label{sec4}

We have argued above hat the study of medium modified parton fragmentation provides 
access to partonic thermalization processes. The remnants of a hard parton are thermalized
to the extent to which they cannot be distinguished any more from the typical degrees of 
freedom in the approximate heat bath provided by the medium. Leading hadrons provide 
only partial and highly biased information about the medium-dependent dynamics which 
drives partonic fragmentation towards thermal equilibrium. There is an obvious danger that 
e.g. leading high-$p_T$ trigger particles bias the fragmentation such that information 
about the thermal properties of the {\it average} partonic fragmentation is
lost. The need to study the unbiased average parton fragmentation pattern provides a 
strong motivation to go beyond single inclusive hadron spectra and to analyze the 
medium-modification of internal jet properties. Other motivations
come from arguments that internal jet structures are more sensitive to
properties of the medium~\cite{Salgado:2003rv} and that they allow us to disentangle
different characteristic properties of the medium (such as strength of collective 
motion and local density~\cite{Armesto:2004pt,Armesto:2004vz}) which are difficult to 
disentangle with single inclusive spectra alone. 

\subsection{Broadening of jet energy and multiplicity distributions}

As discussed in section~\ref{sec2}, theory predicts a one-to-one correspondence between 
the energy degradation of the leading parton (governed by the characteristic gluon energy 
$\omega_c = \frac{1}{2} {\hat{q}} L^2$) and the transverse-momentum broadening of the 
parton shower~\cite{Wiedemann:2000tf,Salgado:2003gb} (governed by ${\hat q} L$). To 
go beyond leading hadron spectra, one can study jet-like near-side particle correlations 
associated to high-$p_T$ trigger particles~\cite{Pal:2003zf,Majumder:2004wh}, particle 
distributions associated to trigger particles, or calorimetrically reconstructed ``true'' jets measurements~\cite{Salgado:2003rv,Armesto:2004pt}. One 
possibility to characterize the transverse phase space distribution 
of a jet is by measuring the fraction $\rho(R)$ of the total jet energy 
$E_t$ deposited within a subcone of radius 
$R = \sqrt{(\Delta \eta)^2 + (\Delta \Phi)^2}$,
\begin{eqnarray}
  \rho_{\rm vac}(R) &=& \frac{1}{N_{\rm jets}} \sum_{\rm jets}
  \frac{E_t(R)}{E_t(R=1)}\, .
  \label{4.1}
\end{eqnarray}
Such measurements have been performed e.g. by the D0 collaboration~\cite{Abbott:1997fc}.
If the BDMPS transport coefficient is very large, an experimentally accessible
broadening of this jet shape may be expected. However, for $\hat{q} = 1 {\rm GeV}^2/fm$,
it was found that the broadening of an $E_T = 100$ GeV jet is negligible~\cite{Salgado:2003rv}.
In this case, the average jet energy fraction inside a small jet cone $R = 0.3$ was reduced by 
$\sim$ 3 \% only. This should be compared to estimates of the background energy 
$E_t^{\rm bg}$ deposited inside the same jet cone which for an event multiplicity 
$dN^{\rm ch}/dy = 2500$ at LHC is of roughly the same size, $E_t^{\rm bg} \sim 100$ GeV.
The example indicates that transverse jet energy distributions, even if deviations
from the vacuum jet shapes can be detected, may be of limited sensitivity for a detailed
characterization of the medium. The physics reason is that the most energetic jet
fragments, which dominate measures of the type (\ref{4.1}), are almost collinear and 
change their angular orientation very little by medium-induced scattering.

\begin{figure}[h]
\begin{center}
\epsfysize=8cm\epsffile{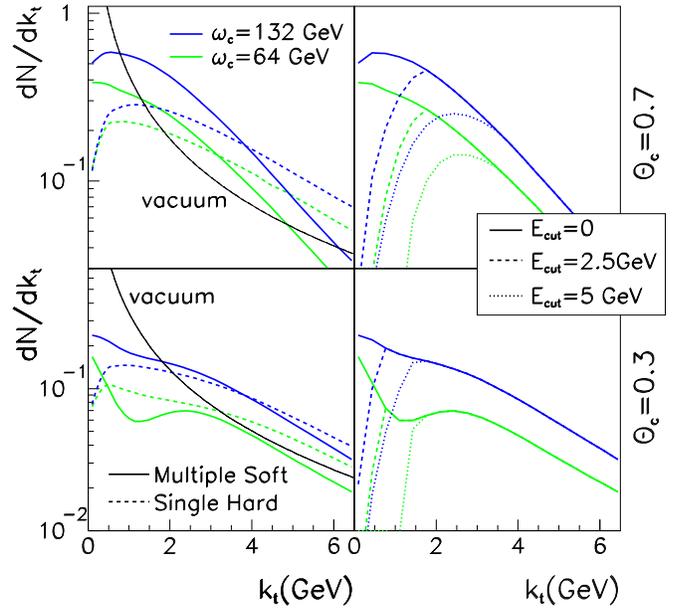}
\caption[a]{The gluon multiplicity distribution
inside a jet cone size $R=\Theta_c$, measured as a 
function of $k_t$ with respect to the jet axis. Removing gluons with
energy smaller than $E_{\rm cut}$ from the distribution (dashed and dotted
lines) does not affect the high-$k_t$ tails. Figure taken from 
\protect\cite{Salgado:2003rv}.}
\label{fig2}
\end{center}
\end{figure}
%

If medium effects do not change the jet energy distribution significantly while the 
energy of leading hadrons decreases strongly, then a significant change in the 
jet multiplicity distributions is unavoidable. It is of particular interest to identify
those multiplicity measurements which are almost insensitive to the high multiplicity 
background in nucleus-nucleus collsions. In Fig.~\ref{fig2}, we show the medium-induced 
additional number of gluons with transverse momentum $k_t = \vert {\bf k}\vert$, produced 
within a subcone of opening angle $\theta_c$, as calculated from (\ref{2.7}), 
see~\cite{Salgado:2003rv}. This distribution is compared to the approximate shape of the 
corresponding distribution in the vacuum.The total partonic jet multiplicity is the sum 
of both components. For realistic values of medium density and in-medium pathlength,
medium effects are seen to increase this multiplicity significantly 
(by a factor $> 2$) in particular in the high-$k_t$ tails. 
Also, the shape and width of the distribution changes
sensitively with the scattering properties of the medium.
Moreover, since gluons must have a minimal energy $\omega > 
k_t/\sin\Theta_c$ to be emitted inside the jet cone, this 
high-$k_t$ tail is unaffected by ``background'' cuts on the 
soft part of the spectrum, see Fig.~\ref{fig2}. 
The insensitivity of the high-$k_t$ tail to the low $E_t$ background 
and its sensitivity to the transverse momentum picked up from the medium
are both based on kinematic grounds and should not depend 
on the details of our calculation.  Thus, the qualitative
conclusions drawn from this study are expected to remain unaffected by the 
uncertainties of our calculation and the effects of subsequent hadronization.
On general grounds, jet multiplicity distributions appear to be more sensitive
to medium properties than jet energy distributions.

\subsection{Interplay of jet quenching and collective flow}

To what extent do jet observables allow us to characterize qualitatively novel properties 
of the produced dense QCD matter which cannot be accessed with leading hadron 
spectra ? The general answer to this question is not known, but first model calculations
indicate that novel information becomes indeed accessible: 

One open issue is that parton energy loss is currently characterized in terms of the BDMPS
transport coefficient which, according to (\ref{3.5}), characterizes the local particle
density of the produced matter. On the other hand, there is strong experimental evidence 
that the produced medium is - if at all - only {\it locally} equilibrated and is thus characterized 
only {\it locally} by its energy density. Measurements of low-$p_T$ inclusive hadron 
spectra~\cite{Adler:2003cb,Adams:2003xp} and their azimuthal asymmetry~\cite{Ackermann:2000tr,Adcox:2002ms,Adler:2003kt} support the picture that different hadron species emerge 
from a common medium which has built up a strong collective velocity field~\cite{Kolb:2003dz}. 
These measurements are broadly consistent with calculations based on ideal 
hydrodynamics~\cite{Kolb:2003dz}, in which the dynamic behavior of 
the produced QCD matter is fully specified by its equation of state 
$p = p(\epsilon,T,\mu_B)$ which enters the energy momentum tensor
\begin{equation}
  T^{\mu\nu}(x) = \left(\epsilon + p \right)\, u^{\mu}\, u^{\nu}\, 
                  - p\, g^{\mu \nu}\, . 
  \label{4.2} 
\end{equation}
%
%
\begin{figure}[h]
\begin{center}
\includegraphics[width=5.3cm,angle=-90]{carlosflows.epsi}
\end{center}
\begin{center}
\includegraphics[width=8.5cm]{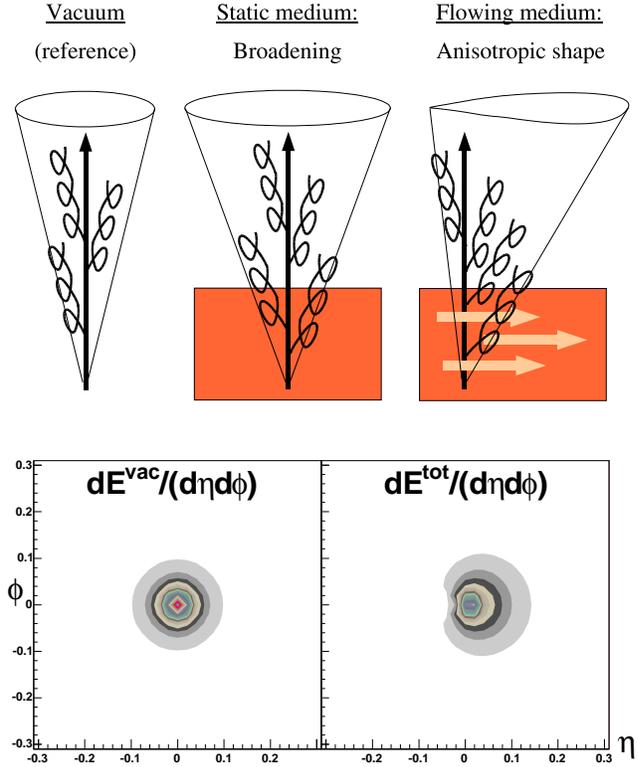}
\end{center}
\vspace{-0.5cm}
\caption{Upper part: sketch of the distortion
of the jet energy distribution in the presence of a medium
with or without collective flow. Lower part: calculated
distortion of the jet energy distribution
in the $\eta \times \phi$-plane for a 100 GeV jet. The right
hand-side is for an 
average medium-induced radiated energy of 23 GeV and equal
contributions from density and flow effects.
Figures taken from~\protect\cite{Armesto:2004pt}.
}\label{fig3}
\end{figure}

Shouldn't parton energy loss be sensitive to $T^{\mu\nu}(x)$ and thus to a combination
of particle density and flow effects ? To quantify the effect of collective flow on parton 
fragmentation, we have extended the formalism of medium-induced parton energy loss 
(\ref{2.7}) to the case of flow-induced azimuthally oriented momentum 
transfer~\cite{Armesto:2004pt,Armesto:2004vz}. One finds 
that flow effects generically result in characteristic asymmetries in the 
$\eta \times \phi$-plane of jet energy distributions and of multiplicity distributions 
associated to high-$p_T$ trigger particles, see Fig.~\ref{fig3}. First experimental
support for this picture comes from the enhanced and broadened rapidity distribution
of hadron production associated to trigger particles~\cite{Wang:2004kf,Armesto:2004pt}. 
Flow effects on the associated initial state radiation may also play a role in accounting 
for these data~\cite{Voloshin:2003ud}. Moreover, model calculations indicate that 
collective flow also contributes to the medium-induced suppression of 
single inclusive high-$p_T$ hadron spectra. In particular, one finds that 
low-$p_T$ elliptic flow can induce a sizeable additional  
contribution to the high-$p_T$ azimuthal asymmetry by selective elimination
of those hard partons which propagate with significant inclination
against the flow field~\cite{Armesto:2004vz}. This reduces at least partially the problem
that models of parton energy loss tend to underpredict the 
large azimuthal asymmetry $v_2$ of high-$p_T$ hadronic spectra
in semi-peripheral Au+Au collisions~\cite{Drees:2003zh,Dainese:2004te}. 
This is a first example of how a detailed dynamical
characterization of the internal jet structure may give access to qualitatively
novel information about the collective properties of the produced matter.
Another recent suggestion is based on the observation that in an ideal fluid, 
energy can be radiated only by sound waves. This could lead to the dramatic 
phenomenon that hadron production associated to high-$p_T$ trigger particles
emerges like a sonic boom preferentially under angles characteristic of the sound
velocity of the produced medium~\cite{Casalderrey-Solana:2004qm}.

In conclusion, the study of jet-like particle correlations and jet measurements
in heavy ion collisions is still at the very beginning. There are strong motivations
to expect that analyzing the interplay of parton fragmentation and medium
properties will provide a laboratory for QCD equilibrium and non-equilibrium
physics which is of unmatched versatility and accuracy.

I thank Nestor Armesto, Andrea Dainese, Alex 
Kovner and Carlos Salgado for the work done together. I am particularly grateful
to Peter Jacobs for many discussions and helpful comments about this manuscript. 


\end{document}